\documentclass[aps,prd,floatfix,twocolumn,notitlepage,superscriptaddress,preprintnumbers,nofootinbib,10pt]{revtex4-1}

\usepackage{graphicx}
\usepackage{amssymb}
\usepackage{amsmath,amssymb}
\usepackage{color}
\usepackage{setspace}
\usepackage{fancyhdr}
\usepackage{multirow}

\definecolor{darkblue}{rgb}{0,0,0.5}
\usepackage[colorlinks,linkcolor=blue,citecolor=blue,urlcolor=blue]{hyperref}

\allowdisplaybreaks

\newcommand\beq{\begin{equation}}
\newcommand\eeq{\end{equation}}

\newcommand\hc{\text{h.c.}}

\def\sol{\rm sol}
\def\atm{\rm atm}

\def\CP{$CP$~}

\usepackage{titlesec}
\begin{document}

\title{Minimal Zee model for lepton ${g-2}$ and $W$-mass shifts}

\author{R. Primulando}
\email{rprimulando@unpar.ac.id}
\affiliation{Center for Theoretical Physics, Department of Physics, Parahyangan Catholic University, Jl. Ciumbuleuit 94, Bandung 40141, Indonesia}

\author{J. Julio} 
\email{julio@brin.go.id}
\affiliation{National Research and Innovation Agency, KST B.\,J.\,Habibie, South Tangerang 15314, Indonesia}

\author{P. Uttayarat}
\email{patipan@g.swu.ac.th}
\affiliation{Department of Physics, Srinakharinwirot University, 114 Sukhumvit 23 Road, Wattana, Bangkok
10110, Thailand}

\begin{abstract}
We present a unique Yukawa structure of the Zee model that can accommodate neutrino oscillation data, solves the muon ${g-2}$ problem, and explains the recent $W$-boson mass measurement. Our Yukawa structure is minimal in the sense that it contains the least possible number of parameters. In this minimal scenario, neutrino masses are quasidegenerate and are compatible with both normal and inverted orderings. The mixing angle $\theta_{23}$ is predicted to lie in the second (first) octant for normal (inverted) ordering. In both cases, the \CP violating phase is close to 3$\pi/2$. The minimal texture also predicts a large branching fraction of the heavy neutral Higgs boson into a pair of electron and muon.
\end{abstract}

\maketitle

\flushbottom
\section{INTRODUCTION}

The massive nature of neutrinos is the key evidence that the Standard Model (SM) is incomplete. For the past 20 years, it has been the only conclusive experimental evidence we have. However, recently two new experimental results have emerged that are in serious tension  with the SM predictions. First, the measurement of the anomalous magnetic dipole moment of the muon by the Fermilab Muon ${g-2}$ Collaboration~\cite{Muong-2:2021ojo} confirms the previous Brookhaven result \cite{Muong-2:2006rrc}. Together they worsen the disagreement with the SM prediction~\cite{Aoyama:2020ynm,Aoyama:2012wk,Czarnecki:2002nt,Gnendiger:2013pva,Davier:2017zfy,Keshavarzi:2018mgv,Colangelo:2018mtw,Hoferichter:2019mqg,Davier:2019can,Keshavarzi:2019abf,Kurz:2014wya,Melnikov:2003xd,Masjuan:2017tvw,Colangelo:2017fiz,Hoferichter:2018kwz,Gerardin:2019vio,Bijnens:2019ghy,Colangelo:2019uex,Blum:2019ugy,Colangelo:2014qya} to a $4.2\sigma$ level. This is commonly referred to as the muon ${g-2}$ problem, and its resolution, barring a major revision in the SM prediction, requires a new physics contribution with $\delta a_\mu = (251\pm59)\times10^{-11}$. Second, the CDF measurement of the $W$-boson mass has pushed the world average value to $m_W = 80.4242\pm0.0087$ GeV~\cite{CDF:2022hxs}, resulting in a 6.1$\sigma$ tension with the SM prediction~\cite{Workman:2022ynf}. We will refer to this as the $W$-mass problem.

The Zee model of neutrino masses~\cite{Zee:1980ai} contains enough ingredients to address those problems.\footnote{For other works that can simultaneously explain the muon {$g-2$} and the $W$-mass problems, see Refs.~\cite{Babu:2022pdn,Han:2022juu,Kawamura:2022uft,Nagao:2022oin,Arcadi:2022dmt,Bhaskar:2022vgk,Baek:2022agi,Zhou:2022cql,Kim:2022hvh,Kim:2022xuo,Dcruz:2022dao,Chowdhury:2022dps,Kim:2022zhj,Chakrabarty:2022voz,Batra:2022pej,Batra:2022arl,Batra:2022org,Chowdhury:2022moc}.} It induces neutrino masses via radiative corrections with TeV-scale physics. It has several Yukawa couplings that, together with such mass scale, can induce a sizable anomalous magnetic moment of the muon. The $W$-mass problem can be solved by large oblique parameters, significantly deviating from zero, thanks to its scalar content. Recently, Ref.~\cite{Chowdhury:2022moc} has demonstrated an example of how to alleviate those problems in the Zee model. However, their solutions  contain a large number of free parameters in the leptonic Yukawa sector, making it rather difficult to test the model quantitatively. 

In its general form, the Zee model is no stranger to such a large number of parameters. Being an extension of the two-Higgs-doublet model with an extra singly charged scalar, it is natural to expect tree-level flavor-violation processes. 
Some authors have tried to cope with these issues by introducing an extra symmetry. The first attempt was done by Wolfenstein \cite{Wolfenstein:1980sy}, who imposed a discrete symmetry such that only one of the Higgs doublets can couple to fermions. As a result, tree-level flavor-changing neutral currents (FCNCs) are forbidden, but  the induced neutrino mass matrix has  vanishing diagonal elements. Such a structure, once compatible with the bimaximal neutrino mixing, has been ruled out by recent solar and KamLAND data~\cite{Super-Kamiokande:2016yck}, which prefer a large but not maximal solar mixing angle. For earlier analyses on this matter, see Refs.~\cite{Petcov:1982en,Smirnov:1994wj,Smirnov:1996bv,Jarlskog:1998uf,Frampton:1999yn,Koide:2001xy,Frampton:2001eu,He:2003ih}.

It should be noted that,  in a more relaxed assumption, in which both Higgs doublets can couple to leptons, the Zee model is still viable  (see, e.g., \cite{Balaji:2001ex,Hasegawa:2003by,He:2003ih,Babu:2013pma,Primulando:2022lpj,Nomura:2019dhw}). Here, tree-level lepton-flavor-violation (LFV) processes are not completely forbidden but their rates can be tamed to be below their respected bounds. This is particularly realized in cases with flavor-dependent symmetries~\cite{Babu:2013pma,Primulando:2022lpj}; in addition, the number of free parameters of the model is  greatly reduced, so that the neutrino mass matrix can be expressed in terms of a few parameters, allowing an interplay between neutrino oscillation data and LFV rates.

A well-defined question can be asked: {\it within the Zee model, how many parameters are actually needed to explain neutrino oscillation data together with muon ${g-2}$ and $W$-mass problems?}   The answer to this question is the objective of this article. We shall systematically search for a minimal leptonic Yukawa structure  that can accommodate neutrino masses and mixing and solve the muon ${g-2}$ and the $W$-mass problems. Any such solutions must also be consistent with existing experimental constraints, in particular, those of  lepton-flavor violations.

\newpage

\section{OVERVIEW OF THE ZEE MODEL}
\label{sec:Zeemodel}
The Zee model is an extension of the two-Higgs-doublet model (2HDM) by a singly charged scalar $\eta^+$.  The model can be most conveniently described in the Higgs basis~\cite{Georgi:1978ri}, where the two Higgs doublets, $H_{1,2}$, are
\begin{equation}
	H_1 = \begin{pmatrix}G^+\\[0.2em] \dfrac{v+h_1+iG}{\sqrt{2}} \end{pmatrix},\quad
	H_2 = \begin{pmatrix}H^+ \\[0.2em] \dfrac{h_2+iA}{\sqrt{2}}\end{pmatrix}.
	\label{eq:doublet}
\end{equation}
Here $v$ is the vacuum expectation value, and $G^+$ and $G$ are the would-be Goldstone bosons. 
In this work, we will assume \CP symmetry in the scalar sector so that $h_{1,2}$ do not mix with $A$. Furthermore, motivated by the current 125-GeV Higgs data, which are consistent with the SM predictions~\cite{Khachatryan:2016vau,CMS:2018uag,ATLAS:2019nkf}, we shall work in the decoupling limit of the 2HDM, so $h_1$ and $h_2$ do not mix. This allows us to identify $h_1$ with the observed 125-GeV Higgs boson, labeled $h$, and $h_2$ with the heavy $CP$-even Higgs boson, henceforth called $H$.\footnote{Typically, the masses of the extra Higgs boson are assumed to be heavier than 125~GeV. However, to the best of our knowledge, current data do allow for the possibility that one or more of these extra Higgs bosons are lighter than 125 GeV.}

In our present work, we choose the basis where charged leptons are diagonal. The leptonic Yukawa interactions responsible for neutrino mass generation are given by
\begin{align}
	\mathcal{L} \supset & \sqrt{2}\frac{(M_{\ell})_{ij}}{v} \bar L_i e_{Rj} H_1 + Y_{ij}\bar L_i e_{Rj}H_2  \nonumber \\ & + f_{ij}L_i^TC \epsilon L_j\eta^+  +\hc,
	\label{eq:yukawa}
\end{align}
where $M_\ell = {\rm diag}\left(m_e,m_\mu,m_\tau\right)$ is the charged lepton mass matrix, $L_i$ and $e_{Ri}$ denote the lepton doublet and singlet of the $i$th generation, respectively, $\epsilon\equiv i\sigma^2$ is an antisymmetric tensor for $SU(2)$ indices contraction, and $C$ is the charge conjugate matrix. The antisymmetric coupling matrix $f$ can be made real by phase rotation, leaving $Y$, therefore, complex. In our current setup, we assume that $H_2$ does not couple with quarks. This implies that quark interactions within this model mimic those of the SM.  Quarks can only couple to $H_1$, inducing their masses according to
\begin{align}
	\mathcal{L} \supset & Y_u \bar Q_{L} u_{R} (\epsilon H_1^*) + Y_d \bar Q_{L} d_{R} H_1 + \hc,\end{align}
where $Y_u$ and $Y_d$ are the corresponding up- and down-type quark Yukawa couplings, respectively.

The presence of Yukawa couplings $f_{ij}$ by itself does not break the lepton number because one can always assign lepton number $-2$ to $\eta^+$. In order to break the lepton number, one  needs to invoke a trilinear coupling, $V \supset \mu H_1 \epsilon H_2 \eta^- +\hc$, which is part of the scalar potential.
Now with both $f$ and the $\mu$ couplings on hand, the lepton number can be broken by two units, leading to a Majorana neutrino mass generation at one-loop level, see  Fig.~\ref{fig:numass}. The trilinear coupling $\mu$ induces a mixing between the two  charged states $H^+$ and $\eta^+$. This mixing, however, is expected to be small because it is proportional to neutrino masses. Thus, to a good approximation, one can treat $H^+$ and $\eta^+$ as (nearly) mass eigenstates. 

The fact that the Zee model is rich in scalars whose masses are about electroweak scale indicates that they may give a significant contribution to the electroweak oblique parameters. It has been shown that such parameters, sensitive to scalar mass splittings, can be responsible to account for the new CDF $W$-mass measurement \cite{Strumia:2022qkt,Lu:2022bgw,Heo:2022dey,Asadi:2022xiy}.

In addition, the Yukawa couplings $f$ and $Y$ also induce lepton anomalous magnetic dipole moments $\delta a_\ell$ and LFV decays. Since $f_{ij}$ couplings (via $\eta^+$ exchange) always give negative $\delta a_\ell$, while data imply that $\delta a_\mu$ has to be positive, we simply diminish their contributions by setting $\eta^+$ to be  heavy and/or $f_{ij}$ to be small; thus, any flavor-violation processes induced by such particle exchange will be negligible too. As for $Y_{ij}$, it is important to note that they cannot be all diagonal, otherwise one cannot reproduce a correct neutrino data fit (see, e.g.,  Ref.~\cite{He:2003ih}). With off-diagonal couplings present, it is natural to expect the occurrences of flavor-violation processes in this model, which also need to be controlled.

\section{PHENOMENOLOGY}
\subsection{Neutrino masses and mixing}
\begin{figure}[b!]
	\begin{center}
		\includegraphics[width=0.38\textwidth]{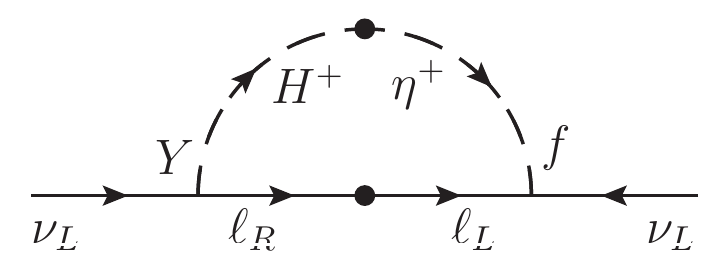}
		\caption{The one-loop diagram generating Majorana neutrino masses. A dot on the internal fermion (scalar) line represents the chiral mass ($\mu$) insertion.}
		\label{fig:numass}
	\end{center}
\end{figure}

The Majorana neutrino mass matrix is induced by Feynman diagram shown in Fig.~\ref{fig:numass}. Since Majorana mass matrix must be symmetric in flavor basis, there is a similar diagram but with internal particles being replaced by their charged conjugates. Altogether they give
\begin{equation}
M_\nu = \kappa(fM_{\ell}Y^\dagger + Y^\ast M_{\ell} f^T), 
\label{eq:nu-massmatrix}
\end{equation}
where $\kappa$ is a constant containing a loop factor and the trilinear coupling $\mu$. Note that in the above equation flavor indices are suppressed.  The flavor structures of the coupling matrices $f$ and $Y$ must be such that they are able to support current neutrino oscillation data,  shown in Table~\ref{tb:oscillation} together with their 1$\sigma$ ranges.

To determine minimal textures leading to correct neutrino data, one should recall that there are five observables that need to be fit; they are the three mixing angles $\theta_{23},\theta_{13},\theta_{12}$, the neutrino mass splitting ratio $R\equiv\Delta m_{\rm sol}^2/\Delta m_{\rm atm}^2$,\footnote{The values of $\Delta m_{\rm sol}^2$ and $\Delta m_{\rm atm}^2$ can be determined by fixing the overall coupling $\kappa$ once the ratio is determined.} and the Jarlskog invariant $J_{CP}$. The two coupling matrices $f$ and $Y$ contain $m$  and $n$ independent, nonvanishing components, respectively. Since one component of each $f$ and $Y$ can be scaled out, effectively  there are $m+2n-3$ independent real parameters that contribute to explaining those five observables. We found that the texture where $(m,n)=(3,2)$ contains the least number of real parameters that can accommodate oscillation data. Other $(m,n)$ textures with equal or smaller number of parameters lead to a neutrino mass matrix with three or more independent zeros, and therefore they are incompatible with neutrino data~\cite{Frampton:2002yf}. Note that, for the aforementioned minimal texture $(m,n)=(3,2)$,  the number of effective parameters is less than the oscillation observables (i.e., 4 vs. 5). This further indicates that at least one of the oscillation parameters can be determined in terms of the others.

\begin{table}[t!]
	\begin{center}
		\caption{Central values and the 1$\sigma$ ranges for neutrino oscillation parameters obtained from the 2021 updated analysis of NuFIT 5.1 (http://www.nu-fit.org), see also \cite{Esteban:2020cvm}. Here and in what follows we use $s^2_{ij}\equiv \sin^2\theta_{ij}$. For $s_{23}^2$, we allow for the possibility that $\theta_{23}$ lies in the shallow minimum.}
		\label{tb:oscillation}
		\begin{tabular}{c|c|c}
			\hline\hline
			Parameters & Normal ordering & Inverted ordering \\
			\hline
			$s^2_{12}$ &$0.304^{+0.012}_{-0.012}$ &$0.304^{+0.013}_{-0.012}$\\
			$s^2_{23}$ &$0.450^{+0.019}_{-0.016}$ &$0.570^{+0.016}_{-0.022}$\\
			$[s^2_{23}$ (shallow)]&$[0.565^{+0.016}_{-0.034}]$ &$[0.455^{+0.021}_{-0.022}]$\\
			$s^2_{13}$ &$0.02246^{+0.00062}_{-0.00062}$ &$0.02241^{+0.00074}_{-0.00062}$\\[.1cm]\hline
			$\Delta m^2_{\sol}/10^{-5}~\text{eV}^2$ &$7.42^{+0.21}_{-0.20}$  &$7.42^{+0.21}_{-0.20}$   \\
			$\Delta m^2_{\atm}/10^{-3}~\text{eV}^2$ &$2.510^{+0.027}_{-0.027}$  &$2.490^{+0.026}_{-0.028}$ \\
			$R$ &$0.0296^{+0.0033}_{-0.0033}$ &$0.0298^{+0.0033}_{-0.0033}$\\[.1cm]\hline
			$J_{CP}$ &$-0.0254^{+0.0115}_{-0.0080}$ &$-0.0330^{+0.0044}_{-0.0011}$\\
			\hline\hline
		\end{tabular}
	\end{center}
\end{table}

Interestingly, with only two nonvanishing $Y_{ij}$, the neutrino mass matrix will have at least two independent zeros, as already discussed in Ref.~\cite{Frampton:2002yf}. Thus, there are only few of such $Y_{ij}$ pairings consistent with neutrino oscillation data and prospectively inducing muon ${g-2}$. In terms of nonvanishing $Y_{ij}$, they are (i) ($Y_{\mu\tau},Y_{\tau e}$), (ii) ($Y_{e\mu},Y_{\tau e}$), (iii) ($Y_{\mu e},Y_{\tau\mu}$), (iv) ($Y_{\mu e},Y_{e\mu}$), and (v) ($Y_{\mu e},Y_{e\tau}$). Note that we do not include in our list pairs of $Y$ couplings that cannot induce $\delta a_\mu$ at all (e.g., $Y_{e\tau},Y_{\tau e}$), although they may give a good fit to neutrino oscillation data. 
 
Further inspection reveals that coupling pairs (i)--(iv) cannot induce enough $\delta a_\mu$ to account for the discrepancy. Pairs (i) and (ii), for example, are restricted by their neutrino flavor structures, while pairs (iii) and (iv) are restricted by flavor constraints.  Coupling pair (i) induces a neutrino mass matrix with vanishing $e$-$e$ and $e$-$\tau$ elements, so it admits the normal mass ordering with $m_1<m_2\ll m_3$. From Eq.~\eqref{eq:nu-massmatrix},  $Y_{\mu\tau}$ always couples with the tau mass, whereas $Y_{\tau e}$ with the electron mass, and for such ordering it is required that $(M_\nu)_{\mu\mu}$ and $(M_\nu)_{\tau\tau}$ be of the same order, and so one would expect that $|Y_{\tau e}/Y_{\mu\tau}|\sim {\cal{O}}(m_\tau/m_e)$. Since $Y_{\mu\tau}$ needs to be of  $\mathcal{O}(1)$ to accommodate $\delta a_\mu$ (see Sec.~\ref{g-2}), this is a clear indication that this pair is not a viable option. In addition, the product of those two couplings is severely constrained by $\mu \to e\gamma$ arising via internal chirality flip of tau mass, which puts $|Y_{\tau e}Y_{\mu\tau}|\lesssim \text{few}\times 10^{-7}$ for $m_{H,A}\sim\mathcal{O}(100)$~GeV.
Similarly, for pair~(ii), neutrino data imply $|Y_{\tau e}/Y_{e\mu}|\sim {\cal O}(m_\mu/m_e)$, which also indicates that a sizable $\delta a_\mu$ cannot be obtained.

As for pair (iii), neutrino fit requires that $|Y_{\mu e}/Y_{\tau\mu}|\sim \mathcal{O}(m_\mu/m_e)$. While it seems that it could induce a correct $\delta a_\mu$, these couplings, however, are constrained by LFV processes, in particular the tree-level $\tau\to \mu\mu e$ and the one-loop $\tau\to e\gamma$. As a result,  the correction to muon ${g-2}$ can only be at most  $115\times 10^{-11}$, which is more than $2\sigma$ away from the global average value. Pair (iv), on the other hand, is free from such LFV decay constraints. However, it suffers from constraint on electric dipole moment of the electron, putting ${\rm Im}(Y_{e\mu}Y_{\mu e})\lesssim10^{-9}$ for $m_{H,A}\sim {\rm few~hundred~GeV}$.  Together with $\delta a_e\sim 10^{-13}$ (see Sec.~\ref{g-2}), this will affect the coupling magnitudes, resulting in a small $\delta a_\mu$. 

It turns out that only (v) can solve the muon ${g-2}$ problem and is consistent with lepton-flavor constraints. This texture leads to a neutrino mass matrix with vanishing $e$-$\tau$ and $\tau$-$\tau$ elements. To get a good fit, one needs $|Y_{\mu e}/Y_{e\tau}| \sim \mathcal{O}(m_\tau/m_e)$. Such coupling hierarchy is central to getting a sizable shift of muon anomalous magnetic moment, and in the same time, suppressing LFV decay rates.

Recall that ${\rm diag}(m_1,m_2,m_3)=U^TM_\nu U$,  with $U$ being the Pontecorvo-Maki-Nakagawa-Sakata matrix.  The two zero conditions can then be solved for $m_1$ and $m_2$, as in \cite{Frampton:2002yf}. Coupled with the fact that $\theta_{23}\simeq \pi/4$ and $\theta_{13}\ll 1$, one can identify that neutrino masses in this scenario are quasidegenerate. The $\theta_{23}$ itself is favored to lie in the second (first) octant for normal (inverted) neutrino mass ordering; either case corresponds to the shallow minimum of the global neutrino fit. Together with $\Delta m_{\rm sol}^2/\Delta m_{\rm atm}^2\ll 1$, the $CP$-violating phase needs to be close to $3\pi/2$ (i.e., nearly maximal). In the upcoming sections, we shall focus our analysis on the context of the ($Y_{\mu e},Y_{e\tau}$) pair. 

\newpage

\subsection{Muon \boldmath{${g-2}$} and flavor-violation constraints}
\label{g-2}
At one-loop level, couplings $Y_{e\tau}$ and $Y_{\mu e}$ induce the following shifts
\begin{align}
	\delta a_e &= \frac{m_e^2}{96\pi^2} \left(\frac{|Y_{e\tau}|^2+|Y_{\mu e}|^2}{m_H^2} + \frac{|Y_{e\tau}|^2+|Y_{\mu e}|^2}{m_A^2} - \frac{|Y_{\mu e}|^2}{m_{H^+}^2}\right),
	\label{dae} \\
	\delta a_\mu &= \frac{m_\mu^2|Y_{\mu e}|^2}{96\pi^2}\left(\frac{1}{m_H^2} + \frac{1}{m_A^2}\right).
	\label{damu}
\end{align}
For exotic Higgs boson masses at a few hundred GeV, one needs $|Y_{\mu e}|\sim\mathcal{O}(1)$ to account for the present discrepancy, which is given at $\delta a_\mu=(251\pm 59)\times10^{-11}$~\cite{Muong-2:2021ojo}.  Such ${\cal O}(1)$ coupling also induces the electron anomalous magnetic moment, so we must ensure that the induced $\delta a_e$ is also consistent with data.

The main issue with the electron magnetic moment is its SM value, which is sensitive to the fine-structure constant. Recently, two measurements, one using a cesium atom~\cite{Parker:2018vye} and the other a rubidium atom~\cite{morel2020}, have  found $\alpha_{\rm Cs}^{-1}=137.035999046(27)$ and $\alpha_{\rm Rb}^{-1}=137.035999206(11)$, which differ by more than $5\sigma$. When translated to the electron magnetic moment, the Cs result suggests a $-2.4\sigma$ discrepancy, while the Rb suggests $+1.6\sigma$, each with respect to the direct measurement~\cite{Hanneke:2008tm}. In our analysis, we treat the two measurements as independent. Specifically, we combine them and infer the SM prediction for the electron magnetic moment \cite{Aoyama:2019ryr}. It is then compared with the direct measurement to get $\delta a_e^{\rm comb}=(2.8\pm3.0)\times 10^{-13}$. Note that the combined $\delta a_e$ is positive, thanks to the smaller error of the Rb experiment compared to that of the Cs.

In the present case, since the neutrino data fit dictates $|Y_{e\tau}|\ll |Y_{\mu e}|$, one can always ignore $Y_{e\tau}$, leading to an interesting relation, namely,
\begin{align}
\delta a_e \simeq \delta a_\mu \left(\frac{m_e^2}{m_\mu^2}\right)\left(1-\frac{m_H^2m_A^2}{(m_H^2+m_A^2)m_{H^+}^2} \right).
\end{align}
This shows that, if $\delta a_\mu$ is found to be of order $10^{-9}$ for $m_H,m_A,m_{H^+}\sim~\text{few~hundred~GeV}$, $\delta a_e$ is predicted to be less than $1\times10^{-13}$, which is in perfect agreement with the $\delta a_e^{\rm comb}$ given above.

The couplings $Y_{e\tau}$ and $Y_{\mu e}$ lead to LFV decays $\tau\to\mu ee$ at tree level and $\tau\to\mu\gamma$ and $\tau\to3\mu$ at one-loop level. Note that the one-loop processes are suppressed by the internal electron mass, so we expect the rates to be well below their bounds. Explicit expressions for LFV observables in the decoupling limit can be found in Ref.~\cite{Primulando:2022lpj}. Here, we note that   amplitudes of these LFV decays are proportional to $|Y_{e\tau}Y_{\mu e}|$. For $Y_{\mu e}\sim\mathcal{O}(1)$ and the heavy Higgs masses around the weak scale, the tree-level LFV process implies $|Y_{e\tau}| \lesssim 10^{-3}$--$10^{-2}$, also consistent with the result from the neutrino data fit.
On the other hand, LFV processes induced by the antisymmetric couplings $f_{ij}$ are negligible because we assume that the scale of $f$ is small and $\eta^+$ is heavy as mentioned at the end of Sec.~\ref{sec:Zeemodel}.

\subsection{\boldmath{$W$}-boson mass}
With radiative corrections, the $W$ mass is expressed by the following relation \cite{Sirlin:1980nh}:
\begin{align}
m_W^2= \frac{\pi \alpha}{\sqrt{2}G_Fs_W^2}(1+\Delta r),
\label{delta-r}
\end{align}
where $\alpha=1/137.036$ is the QED fine-structure constant, $G_F$ is the Fermi constant determined from the muon decay, $s_W^2\equiv1-m_W^2/m_Z^2$, with $m_Z=91.1876$ GeV being the $Z$-boson mass, represents the weak mixing angle, and $\Delta r$ is a parameter that contains loop corrections to the tree-level one. 

The parameter $\Delta r$ has two parts: one is due to electroweak oblique parameters, and the other due to vertex and box contributions to the muon  decay~\cite{Lopez-Val:2012uou}. In the present scenario, the new physics effects to the second parts are dominated by diagrams modifying the $W$-$\mu$-$\nu_\mu$ vertex via $g\to g(1+\delta g_\mu)$, with $g$ being the $SU(2)$ gauge coupling, similar to the one discussed in \cite{Abe:2017jqo}. The shift is given by
\begin{align}
\delta g_\mu = \frac{|Y_{\mu e}|^2}{32\pi^2} \left[1-\xi(m_H^2,m_{H^+}^2)-\xi(m_A^2,m_{H^+}^2) \right].
\end{align}
with
\begin{align}
\xi(x,y) = \frac{x+y}{4(x-y)}\ln \frac{x}{y}.
\end{align}
Such a gauge coupling correction might not be negligible, particularly for large $Y_{\mu e}$. However, its value is restricted by the measurements of lepton-flavor universality $g_\mu/g_e=1.0018\pm0.0014$ \cite{Pich:2013lsa}, with $g_\ell$ being the corresponding $W$-$\ell$-$\nu_\ell$ coupling.  A similar correction also exists for $g_e$, but it is deemed irrelevant because the coupling $Y_{e\tau}$ is  small as required by neutrino data. 

To calculate the $W$ mass, it is more beneficial to rewrite Eq.~\eqref{delta-r} such that all SM contributions are subtracted (i.e., absorbed into $m_W^2|_{\rm SM}$)~\cite{Grimus:2008nb}. That is,
\begin{align}
m_W^2 = & \left. m_W^2 \right|_{\rm SM}  \left( 1+ \frac{s_W^2}{1-2s_W^2} \Delta r' \right),
\label{eq:Wmass}
\end{align}
where $m_W^2|_{\rm SM}=(80.357~\text{GeV})^2$ is the $W$-mass squared predicted by the SM and
\begin{align}
\Delta r' = \frac{\alpha}{s_W^2}\left( -\frac{1}{2}S + (1-s_W^2) T \right)-2\delta g_\mu,
\end{align}
is the new physics contributions. 
We have neglected the $U$ parameter contribution because we found its value  small compared to the $S$ and the $T$ parameters. 
In the limit where $\eta^+$ is decoupled with other scalars, those parameters are given by~\cite{Grimus:2008nb}
\begin{widetext}
\begin{align}
	S &= \frac{1}{24\pi}\left[(1-2s_W^2)^2G(m_{H^+}^2,m_{H^+}^2,m_Z^2) + G(m_{H}^2,m_{A}^2,m_Z^2) - \ln\frac{m_{H^+}^2}{m_H^2} - \ln\frac{m_{H^+}^2}{m_A^2}\right],
	\label{eq:S}\\
	T &= \frac{1}{16\pi^2\alpha v^2}\left[F(m_{H^+}^2,m_H^2) + F(m_{H^+}^2,m_A^2)  - F(m_A^2,m_H^2) \right],
	\label{eq:T}
\end{align}
\end{widetext}
where 
\begin{align}
	F(x,y) = \frac{x+y}{2} - \frac{xy}{x-y}\ln\frac{x}{y}.
\end{align}
The explicit form of the $G(x,y,z)$ function can be found in Appendix C of Ref.~\cite{Grimus:2008nb}. We only note that $G(x,y,z)$ is symmetric in its first two arguments.

The $T$ parameter is generally larger than the $S$ parameter, so the former is supposed to  play a greater role in the $W$-mass shift. However, it is worth noting that in our setup there is an interplay between oblique parameters and the vertex correction in getting the correct $W$ mass. The $\delta g_\mu$, despite always inducing a positive shift in the $W$ mass, is not allowed to be arbitrarily large because its value is constrained by the lepton-flavor universality measurements, taken in our calculation to be within $2\sigma$. For that reason, a sizable, nonvanishing $T$ in the range of
$T\in(0.06,0.2)$ is still needed to account for the new average of the $W$ mass, $m_W=80.4242\pm0.0087$ GeV~\cite{CDF:2022hxs}. Since the $T$ parameter vanishes when either one of the neutral Higgs bosons is degenerate with the charged Higgs, it is also necessary that the masses of the extra Higgs bosons be sufficiently split.

It should be noted that the extra Higgs boson masses are constrained by collider searches as well. In our minimal scenario, they do  not couple to quarks. As a result, the $H$ and $A$ can be pair produced via Drell-Yan processes, leading to multilepton signatures. Similarly, the pair-produced charged Higgs would lead to a dilepton and missing energy signatures. This process has been actively searched for at the LHC. The current experimental limit, assuming $H^+H^-\to\mu^+\mu^-/e^+e^-+\slash\hspace{-7pt}E_T$, puts $m_{H^+}\gtrsim 550$ GeV~\cite{CMS:2018eqb,ATLAS:2019lff}.

\section{EXPLICIT EXAMPLE}
\label{sec:benchmark}
As an illustrative example, let us consider two benchmarks for the Yukawa couplings as follows
\begin{itemize}
\item Benchmark B1 (normal ordering)
\begin{equation}
\begin{aligned}
	\frac{f_{e\mu}}{f_{\mu\tau}} &= 1.082,\quad
	\frac{f_{e\tau}}{f_{\mu\tau}} = 8.285,\\
	\frac{Y_{\mu e}}{Y_{e\tau}} &=  (-7.660 + 1.941i)\times 10^3, \quad 
\end{aligned}
\end{equation} 
\item Benchmark B2 (inverted ordering)
\begin{equation}
\begin{aligned}
	\frac{f_{e\mu}}{f_{\mu\tau}} &= 1.240,\quad
	\frac{f_{e\tau}}{f_{\mu\tau}} = 11.13,\\
	\frac{Y_{\mu e}}{Y_{e\tau}} &= (6.191+0.956i)\times 10^3, \quad %
\end{aligned}
\end{equation} 
\end{itemize}
Both benchmark values give a good fit to all five neutrino oscillation parameters, as shown in  Table~\ref{tb:nupheno}.  Once we determine the mass splitting ratio, the overall neutrino mass can be deduced. From here, we find $\sum m_\nu = 0.196\,(0.229)$~eV for the B1 (B2) case, which is consistent with the bound from the cosmic microwave background and the baryonic acoustic oscillation measurements, setting $\sum m_\nu\le 0.515$ eV~\cite{DiValentino:2019dzu}.  It is not surprising that neutrino mass sums in both B1 and B2 are  comparable, thanks to the quasidegeneracy property of neutrino masses in this scenario. 

Given all coupling ratios above, we consider a set of scalar masses $(m_H,m_A,m_{H^+})=(375,520,550)$ GeV and $Y_{\mu e}=3.4$. With these values, the two benchmarks give $m_W=80.4295$ GeV and $\delta a_\mu=147\times 10^{-11}$, which are within $2\sigma$ of the respective experimental values.  As anticipated, $\delta a_e=0.24\times 10^{-13}$ for both B1 and B2.
The branching ratios for the tree-level LFV decays $\tau\to\mu ee$ are found to be about 2--3 orders of magnitude below the current experimental bounds.  The branching ratios for loop-induced processes, i.e., $\tau\to\mu\gamma$ and $\tau\to3\mu$, owing to the electron mass suppression, are found  to be several orders of magnitude below their experimental bounds.

\begin{table}[t!]
\begin{center}
	\caption{Neutrino oscillation parameters and the sum of neutrino masses in the benchmark scenarios.}
	\label{tb:nupheno}
\begin{tabular}{c|c|c|c|c|c|c}
\hline\hline
Benchmark &$s^2_{12}$ & $s^2_{23}$ & $s^2_{13}$ & $R$ & $J_{CP}$& $\sum m_\nu$ (eV) \\
\hline
B1 & 0.305 & 0.567 & 0.0220 & 0.0298 & -0.0331 &0.196 \\
B2 & 0.299 & 0.442 & 0.0220 & 0.0296 & -0.0330 &0.229 \\
\hline\hline
\end{tabular}
\end{center}
\end{table}

\section{CONCLUSION AND DISCUSSION}
\label{sec:conclusion}
We have identified the minimal Yukawa structure of the Zee model that can accommodate neutrino oscillation data, muon ${g-2}$, and $W$ mass, and are consistent with LFV constraints. Such a structure consists of the $f$ coupling matrix with all three independent components and the $Y$ coupling matrix with only $Y_{e\tau}$ and $Y_{\mu e}$ being nonvanishing. 

Our minimal structure features quasidegenerate neutrino masses,  accommodating both normal and inverted orderings. The mixing angle $\theta_{23}$ lies in the second and first octants for normal and inverted mass orderings, respectively. To guarantee a successful fit, one also needs to have  $|Y_{\mu e}/Y_{e\tau}|\sim m_\tau/m_e$. Such a large ratio plays an important role in getting the desired value of $\delta a_\mu\sim150\times 10^{-11}$, while keeping $\delta a_e\lesssim 10^{-13}$ consistent with data. Furthermore, rates for lepton-flavor-violation processes, such as $\tau \to \mu ee$, $\tau \to e\gamma$, and $\tau \to 3\mu$, appear to be well below their respective bounds, partly due to the large coupling hierarchy and electron mass suppression in $\tau \to \mu$ transitions.

The flavor structure of the neutrino mass matrix in this scenario only supports a shallow minimum of $\theta_{23}$.   Should it be ruled out, one needs to go to the next-to-minimal Yukawa structure. For example, the (2,3) Yukawa texture with nonzero $f_{e\tau}$, $f_{\mu\tau}$, $Y_{e\tau}$, $Y_{\mu e}$ and $Y_{\tau\tau}$ can accommodate the inverted mass ordering with $\theta_{23}$ in the second octant. 

In our minimal scenario, we have $Y_{\mu e}\sim{\cal O}(1)$. This would lead to large $H/A\to e^\pm\mu^\mp$ decays, which could be searched for at the LHC. However, $H/A$ cannot be singly produced in our scenario. They can be pair produced via Drell-Yan processes, leading to multilepton signatures. We leave a careful collider study of such processes for possible future work.

\section*{ACKNOWLEDGMENTS}
The work of RP was supported by the Parahyangan Catholic University under Grant No. III/LPPM/2023-02/32-P.
The work of PU was supported in part by Thailand National Science, Research and Innovation Fund (NSRF) via Program Management Unit for Human Resources \& Institutional Development, Research and Innovation (PMU-B) under Grant No. B05F650021.
PU also acknowledges the supporting computing infrastructure provided by NSTDA, CU, CUAASC (Chulalongkorn Academic Advancement into its 2nd Century Project, Thailand).

\bibliographystyle{apsrev4-1}
\bibliography{reference} 

\end{document}